# Fabrication of Atomically Precise Nanopores in Hexagonal Boron Nitride

S. Matt Gilbert[‡,1,2,3], Gabriel Dunn[‡,1,2,3], Thang Pham[1,2,3], Brian Shevitski[1,2,3,4], Edgar Dimitrov[1,3], Shaul Aloni[4] and Alex Zettl[*,1,2,3].

[1] Department of Physics, University of California at Berkeley, Berkeley, CA 94720, USA
[2] Materials Sciences Division, Lawrence Berkeley National Laboratory, Berkeley, CA 94720, USA
[3] Kavli Energy NanoScience Institute at the University of California, Berkeley and the Lawrence Berkeley National Laboratory, Berkeley, CA 94720, USA
[4] Molecular Foundry, Lawrence Berkeley National Laboratory, Berkeley, CA 94720, USA

KEYWORDS (Boron Nitride, nanopore, DNA sequencing, water desalination").

**ABSTRACT:** We demonstrate the fabrication of individual nanopores in hexagonal boron nitride (hBN) with atomically precise control of the pore size. Previous methods of pore production in other 2D materials create pores of irregular geometry with imprecise diameters. By taking advantage of the preferential growth of boron vacancies in hBN under electron beam irradiation, we are able to observe the pore growth via transmission electron microscopy, and terminate the process when the pore has reached its desired size. Careful control of beam conditions allows us to nucleate and grow individual triangular and hexagonal pores with diameters ranging from subnanometer to 6nm over a large area of suspended hBN using a conventional TEM. These nanopores could find application in molecular sensing, DNA sequencing, water desalination, and molecular separation. Furthermore, the chemical edge-groups along the hBN pores can be made entirely nitrogen terminated or faceted with boron-terminated edges, opening avenues for tailored functionalization and extending the applications of these hBN nanopores.

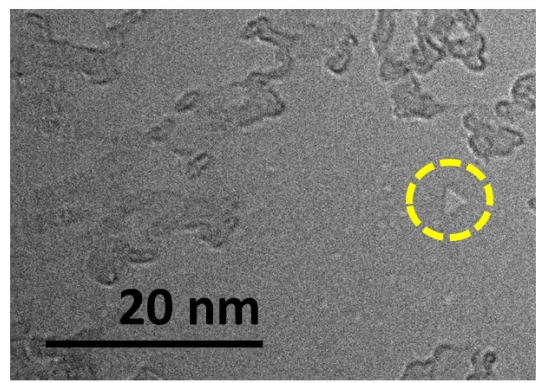

Nanoporous materials have received a great deal of attention due to a range of applications: molecular sensing, DNA sequencing, water desalination, and molecular separation to name a few[1–4]. Nanopores in 2D materials such as graphene[5–8], hexagonal boron nitride (hBN)[9], and molybdenum disulfide (MoS$_2$)[10] have gained particular attention. Limiting the pore's channel length to a single or few atoms makes for more sensitive sensors or more energy efficient filters. Additionally, the crystalline nature of layered materials creates pores in which the chemical groups at the pore's edge differ from those of the rest of the sheet, allowing for chemical functionalization to improve performance.

The performance in each of the applications mentioned above relies critically on precise control of pore size. In molecular sensing, more precise pore sizes result in a larger signal as a target molecule more fully blocks the channel, leading to improved sensing accuracy[11,12]. DNA sequencing would not only benefit from this improved signal, but more precise pore sizes would also help slow the passage of DNA through the pore – a common problem facing many solid state nanopore devices aimed at sequencing[13,14]. Atomically precise pore size would allow for molecular sieves and gas separation systems capable of isolating smaller molecules and separating species with atomic scale differences in size and shape[4,15]. While other techniques have had success in creating nanopores in 2D materials, these previous techniques have lacked the atomic precision[12,16].

In this work, we describe the fabrication of individual nanopores in h-BN with atomically precise control of pore size. As previously demonstrated[17], under transmission electron microscopy (TEM) electron beam irradiation, preferential ejection of boron atoms due to lower knock-on threshold compared to nitrogen leads to boron monovacancy formation. After that the under-coordinated atoms at the defect edges are then more prone to be sputtered either by zipper-like (atoms by atoms) or chain-like (a bundle of atoms at a time) mechanism resulting in the steady growth of triangular pores[18–23]. Here we show that by careful control of electron beam conditions, the density of monovacancy formation and pore growth rate can be controlled independently, allowing us to create individual nanopores over a large area, with sizes precisely controlled by the length of electron beam exposure.

As has been demonstrated previously, triangular defects in hBN caused by electron beam irradiation display solely nitrogen terminated end-groups[24]. Previous methods of creating graphene nanopores with electron beam irradiation can result in extremely reactive and diverse end-group chemistries that are sensitive to oxidation[25,26]. In contrast, the nitrogen terminated ends of the hBN nanopores produced as described above are resistant to oxidation[9] – allowing for easier cleaning of the nanopore – and the uniformity of the end-groups opens up a clear route to edge functionalization around the entire pore.

**Sample Preparation.** The hBN for this work is prepared by chemical vapor deposition (CVD) on copper foil using an ammonia-borane precursor[27,28]. The growth process is described in supporting information (SI 1). The resulting hBN is few-layer, having between 3-5 layers.

Commercially available $SiN_x$ membrane TEM chips are used to support the hBN. Prior to mounting, a single 20-100nm hole is milled in the $SiN_x$ membrane using either extended exposure with a condensed electron beam in TEM, or a helium ion beam. hBN is then transferred to the $SiN_x$ chip via a polymer assisted method, as described previously. After transfer, the chip is annealed in at 350 °C in a hydrogen environment to remove the polymer. Imaging and nanopore formation are performed in a JEOL 2010 TEM operating at 80 kV, while high resolution images are obtained at the National Center for Electron Microscopy's TEAM 0.5 aberration corrected microscope at 80kV.

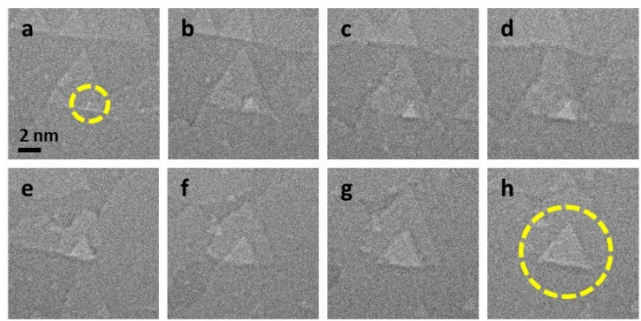

**Figure 2.** Time series showing the quantized growth of a triangular nanopore in hBN. (a) monovacancy formed in bottom hBN sheet, circled in yellow. (b)-(h) Metastable quantized growth of nanopore. Shuttering the electron beam irradiation causes the pore growth to cease.

**Preparation of Atomically Precise Triangular Nanopores.** As demonstrated previously, electron beam irradiation of hBN commonly results in the formation of triangular vacancies, particularly under 80 kV irradiation at room temperature[18,23,29]. This process is due to electron knock-on effects and/or selective chemical etching due to gasses present in the TEM column, either of which preferentially ejects boron atoms and preserves nitrogen zig-zag edge termination[18]. After the formation of a boron monovacancy, the neighboring atoms become more loosely bound and are ejected while continuing to hold a nitrogen terminated zig-zag edge[18,22,23]. Because of this process, the growth of hBN nanopores under electron beam iradiation are metastable at quantized pore sizes, as shown in Figure 1.

| Index | Image | Area |
|---|---|---|
| N = 0 | | $A = 0$ |
| N = 1 | | $A = (2a_o)^2 \frac{3\sqrt{3}}{4}$ <br> $= 0.11 nm^2$ |
| N = 2 | | $A = (3a_o)^2 \frac{3\sqrt{3}}{4}$ <br> $= 0.25 nm^2$ |
| N = n | | $A = ((n+1)a_o)^2 \frac{\sqrt{3}}{4}$ <br> $= (n+1)^2 * 0.027 nm^2$ |

**Figure 1.** For each quantized triangular pore size, an image shows the atomic configuration of the pore and the resultant pore area. Nitrogen and boron atoms are depicted in blue and gold respectively. The spacing between neighboring boron and nitrogen atoms, $a_o$, is 1.45Å.

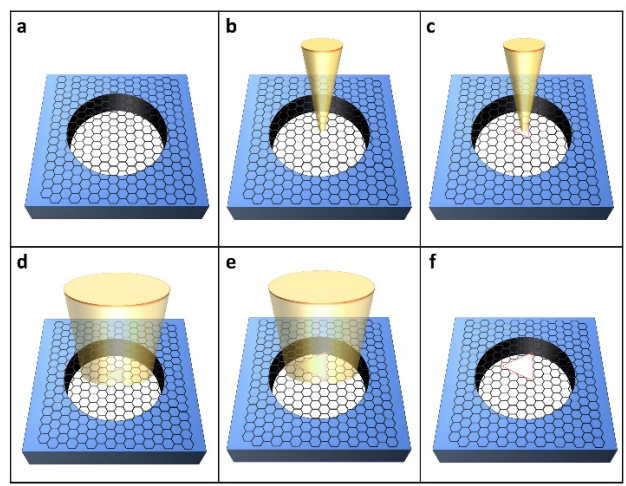

**Figure 3.** Schematic of pore growth process. (a) hBN is suspended over a milled hole in a SiN membrane. (b)-(c) TEM beam is condensed over a small area to nucleate monovacancies. (d)-(e) Beam is spread and spot size reduced to decrease intensity, favoring pore growth over nucleation. (f) Resulting triangular nanopore(s).

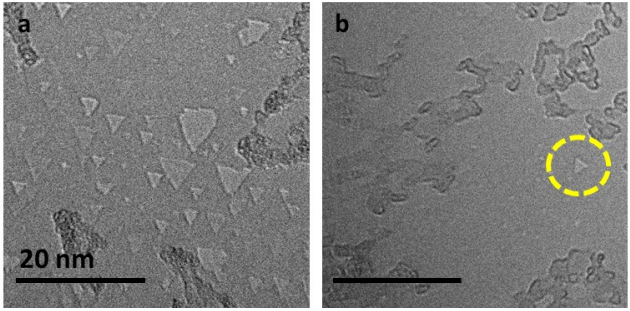

**Figure 4.** Control of monovacancy formation vs pore growth. Condensed beam leads to a number of pores being inititated. Diffuse beam allows for pore growth while limiting new pores being initiated. (a) Initial long exposure to condensed beam leads to a number of pores being initiated. (b) Shortening initial exposure time allows for individual nanopores to be formed. Triangular nanopore circled in yellow.

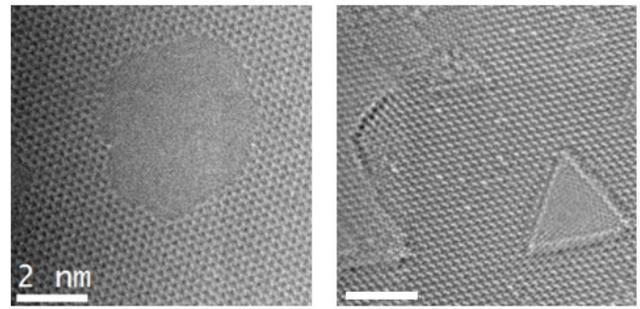

**Figure 5.** (a) Graphene vs (b) hBN nanopores with same technique. Note the irregular pore geometry of (a) vs the pristine triangular geometry of (b). This image was captured using TEAM 0.5 at 80 keV.

By utilizing this vacancy growth mechanism, we can produce nanopores of atomically precise size and shape by monitoring the vacancy growth during irradiation and terminating it by shuttering the electron beam as shown in Figure 2. This allows for reproducible pores of identical shape and size, in stark contrast to methods that rely solely on a condensed electron or ion beam to mill holes in 2D materials. Furthermore, we demonstrate the ability to create single, isolated nanopores over an extended area of suspended hBN.

As illustrated in Figure 3, we first use a condensed electron beam to carefully locally mill through our 3-5 layer hBN over a small area (<5-10nm) for a single pore or large(20-100nm) area for multiple pores for several seconds to several minutes until the first vacancy forms through the entire membrane. We check periodically to see if the entire hBN layer has been pierced by camera contrast.

By controlling the beam conditions (beam size and beam current density), we can favor either vacancy formation or growth. While condensing the beam, vacancies form more readily, allowing us to mill through several layers and create the seeds for our nanopores. As shown in Figure 4, extended initial irradiation with a condensed beam leads to multiple pore sites, whereas brief irradiation leads to individual nanopore sites. By spreading the electron beam, we have found that pore growth is favored over formation, allowing for the isolated growth of individual or few nanopores (Figure 2). Furthermore, by only locally milling through the few layered hBN, additional vacancies that form away from the pore when the beam is spread form only in the top layers.

**Extension and comparison to graphene nanopores.** While this method of vacancy seeding and pore growth can be extended to other 2D materials, hBN nanopores prepared in this manner hold key advantages. When a similar method is used on graphene, the pore area can be precisely controlled by shuttering the beam, but the pore geometry is highly irregular (Figure 5a) as compared to hBN (Figure 5b). Initiating pore growth is also significantly more difficult, requiring a much higher energy or beam current and a fully condensed beam which yields larger starting vacancies. The insulating nature of hBN may be a further advantage over graphene for reducing noise in transmembrane current measurements and in instances where integrating patterned electronics on the substrate is necessary such as in nanoribbon and nanowire or plasmonic nanoantenna detection of nanopores[30–33].

Additionally, the graphene nanopore's edges react readily with air creating a diversity of carboxyl, hydroxyl and epoxide groups when removed from the TEM chamber[25,26]. In comparison, the hBN nanopore edges prepared in this manner have been shown to be entirely nitrogen terminated[17,18]. This makes the hBN nanopores more stable, less susceptible to oxidative reactions – making them easier to clean – and creates a clear path to nanopore functionalization.

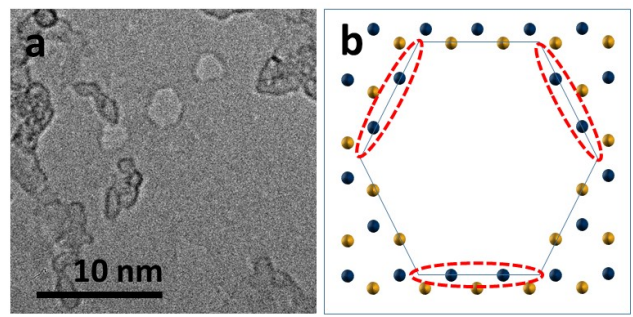

**Figure 6.** (a) Hexagonal hBN pore created by condensing the beam at higher spot sizes to achieve higher current density. (b) Alternating boron and nitrogen facets as demonstrated in Ref. 22. Boron and Nitrogen represented in blue and gold respectively. Boron terminated facets are circled in red.

**Nanopore edge-groups and functionalization.** The functionalization of nanopore edge groups can greatly further their applications. Molecular dynamics simulations have suggested that pore functionalization could be used for selective ion transport through the pore[34] – useful for molecular separation and water desalination. Functionalizing nanopores with specific binding sites for analytes has been demonstrated as a means of greatly improving molecular sensing[35]. Functionalization has also been proposed as a means of slowing the transport of DNA through nanopores for solid state sequencing[36].

Adding to the versatility of these hBN nanopores, we have shown previously that by increasing the temperature of the substrate during electron beam irradiation, the preference for nitrogen-terminated end groups over boron-terminated can be shifted, generating hexagonal pores instead of triangular ones[23]. Here we show that these pores can also be formed at increased beam current densities at 80 kV at room temperature (Figure 6). By condensing the beam at a higher spot size, hexagonal vacancies and nanopores form at the highest current density region of the beam. Hexagonal pores have previously been shown to have facets of alternating boron and nitride termination. This creates a diversity of functionalization approaches and strategies that could further broaden the application of hBN nanopores.

**Conclusion.** In summary, we have developed a method for fabricating individual hBN nanopores with pore sizes that can be controlled with atomic precision. By careful control of beam conditions, electron beam irradiation can be used to preferentially initiate pore creation or pore growth. The resultant pore edges for triangular pores are entirely nitrogen terminated, making the pores stable, largely chemically inert, and conducive to functionalization. These hBN nanopores can find application in DNA sequencing and molecular sensing, where smaller pores and more precise end-group functionalization would increase sensitivity and performance respectively; or in water desalination, where better tailored pore sizes and functionalization could improve performance and efficiency; and in molecular separation, where precise pore size and end group control would allow for better discernment between like chemical species.

## ASSOCIATED CONTENT

**Supporting Information**.
Details of hBN growth and preparation (PDF).

This material is available free of charge via the Internet at http://pubs.acs.org.

## AUTHOR INFORMATION

**Corresponding Author**

* email azettl@berkeley.edu

**Author Contributions**

‡These authors contributed equally to this work.

**Funding Sources**

Any funds used to support the research of the manuscript should be placed here (per journal style).

**Notes**

Any additional relevant notes should be placed here.

## ACKNOWLEDGMENT

This work was supported in part by the Director, Office of Science, Office of Basic Energy Sciences, Materials Sciences and Engineering Division, of the U.S. Department of Energy under Contract No. DE-AC02-05-CH11231, within the sp2-Bonded Materials Program (KC2207) which provided for TEM characterization, and within the Nanomachines Program (KC1203) which provided for methodology for membrane suspension; by the Department of Defense, Defense Threat Reduction Agency under Grant No. HDTRA1-15-1-0036, which provided for He-beam irradiation (the content of the information does not necessarily reflect the position or the policy of the Federal Government, and no official endorsement should be inferred); and by the National Science Foundation under Grant No. DMR-1206512, which provided for additional structural characterization of the pristine membranes. Work at the molecular foundry was supported by the office of science, office of basic energy sciences, of the us department of energy under contract no. DE-AC02-05CH11231. SMG and BS acknowledge support from the National Science Foundation Graduate Fellowship Program.